# A constitutive model for unsaturated cemented soils under cyclic loading

C. Yang [1], Y.J. Cui [2,3], J.M. Pereira [2,3], M.S. Huang [1]

1. Department of Geotechnical Engineering, Tongji University, Shanghai, China
2. Ecole Nationale des Ponts et Chaussées (ENPC) – Navier-CERMES, 6-8 Avenue Blaise Pascal, Cité Descartes, Champs-sur-Marne, F-77455 Marne-la-Vallée Cedex 2, France
3. Université Paris-Est, UR Navier, Marne-la-Vallée, France

**Corresponding author**

Prof. Yu-Jun CUI
Ecole Nationale des Ponts et Chaussées, CERMES
6-8 Avenue Blaise Pascal, Cité Descartes, Champs-sur-Marne
F-77455 MARNE-LA-VALLEE CEDEX 2
France

Email: cui@cermes.enpc.fr
Phone: +33 1 64 15 35 50
Fax: +33 1 64 15 35 62

**Abstract:** On the basis of plastic bounding surface model, the damage theory for structured soils and unsaturated soil mechanics, an elastoplastic model for unsaturated loessic soils under cyclic loading has been elaborated. Firstly, the description of bond degradation in a damage framework is given, linking the damage of soil's structure to the accumulated strain. The Barcelona Basic Model (BBM) was considered for the suction effects. The elastoplastic model is then integrated into a bounding surface plasticity framework in order to model strain accumulation along cyclic loading, even under small stress levels.

The validation of the proposed model is conducted by comparing its predictions with the experimental results from multi-level cyclic triaxial tests performed on a natural loess sampled beside the Northern French railway for high speed train and about 140 km far from Paris. The comparisons show the capabilities of the model to describe the behaviour of unsaturated cemented soils under cyclic loading.

## 1 Introduction

The French high-speed railway line between Paris and Lille (TGV Nord) crosses a widespread range of loess deposits. During the very rainy seasons between Winter 2001 and Spring 2002, instability problems caused by the formation of sinkholes were observed near the railway foundation, with depths up to 7m. First laboratory cyclic tests have showed that this instability would relate to the cyclic behaviour of the involved loessic soil, an aeolian deposited sediment. As this loessic soil contains a significant fraction of carbonates (16%), special attention should be paid to the effect of cementation in the analysis of test results as well as in constitutive modelling. Note that in unsaturated state, the clay fraction contained in the soil (about 16%) can also play a cementation role, even though its cementation level is variable, as a function of the degree of saturation of the soil. This is a particular point of unsaturated fine-grained soils when considering the cementation effects.

Natural cemented soils are often named structured soils because they show a structural strength, i.e. additional strength induced by the specific arrangement of solid grains and the cementation between the solid particles. However, these bonds may present a fragile behaviour and be damaged under mechanical loading, particularly under cyclic loading. Research on constitutive models and damage theory for structured soils has been one of the important topics in the field of soil mechanics in recent years.

Several authors have studied the effects of bond damage of structured soils from a theoretical point of view. Among others, the works of Burghignoli *et al.* (1998), Sharma & Fahey (2003a, b) consider, in terms of consequences of damage, the possibility of changes in the size and shape of the elastic domain (yield surface), and at the same time, the possibility of decrease in the overall soil stiffness. After the complete damage of bond, it is generally accepted that initially structured soil will tend to the same critical state as the equivalent unstructured soil does (see Gens & Nova, 1993; Chai *et al.*, 2005 for instance). In order to quantify the bond damage process, Baudet & Stallebrass (2004) defined a sensitivity coefficient to represent structure and its degradation for all types of loadings. Vaunat & Gens (2003) proposed a coupled formulation between the elastic strains of bond and the total elastic strains of soil and incorporated it into the modified Cam-Clay Model. On the basis of elastoplasticity, Carol *et al.* (2001) established a set of damage theory to delve into the damage evolution of both isotropic and anisotropic materials systematically.

As one efficient tool to simulate the mechanical behaviour of soils under cyclic loading, the bounding surface theory has been extensively applied to various types of soils, especially to clay and sand (Dafalias & Herrmann (1982), Zienkiewicz *et al*. (1985), Pastor *et al*. (1985), Khalili *et al.* (2004)). Chai *et al.* (2005) employed the bounding surface theory to describe the mechanical behaviour of saturated loess under cyclic loading. However, to the authors' knowledge, there has been no publication about the constitutive model of unsaturated structured loess under cyclic loading.

As far as the unsaturated aspects are concerned, intensive research has been made over recent year. Important contributions in the field of constitutive modelling of partially saturated soils have shown that an appropriate framework needs the use of two independent state variables. Net stress and matric suction are often used (see, among others, Alonso *et al*., 1990; Wheeler & Sivakumar, 1995). Note that other choices are possible (see for instance Pereira *et al*. (2005)). In terms of cementation effects, Leroueil & Barbosa (2000) reported that suction increases lead to higher strength and stiffness of both soil matrix and bond. Garitte *et al.* (2006) extended the Barcelona Basic Model (BBM, Alonso *et al*., 1990) starting from the contribution of Vaunat & Gens (2003), thus leading to a model for structured soils taking into account unsaturated states. An energy threshold from which bond damage effectively occurs has been introduced in this extension of BBM. It corresponds to the amount of elastic energy that the bond material is able to store without damage.

This paper aims at developing an elastoplastic model with damage for unsaturated structured soils under cyclic loading within the framework of bounding surface theory. Laboratory experiments are simulated in order to validate the proposed model.

## 2 Geotechnical properties of the studied loess

The soil studied is taken from Northern France, 140 km from Paris along TGV line, at a distance of 25 m from the railway and a depth of 2.2 m. Intact blocs have been sampled. Laboratory identification showed that this loess is a typical homogeneous yellowish-grey, porous calcareous loess (calcium carbonate, $CaCO_3$, up to 16%). It has a low plasticity index (PI = 6), low dry density ($\rho_d = 1.39\,\text{Mg/m}^3$), low degree of saturation ($S_r = 53\%$), low clay fraction (% < 2µm = 16). The grain size distribution curve is depicted in Figure 1.

X-ray diffractometry shows that the sample is mainly composed of quartz. Analysis on the clay fraction (<2 µm) indicates that it involves kaolinite, illite and interstratified illite-smectite.

The microstructure of the soil was observed at Scanning Electron Microscope (SEM) and is shown in Figure 2. In Figure 2-a, a large aggregate of about 200 µm in diameter is observed together with large inter-aggregates pores (up to 30 µm in diameter). Aggregates are made up of silt grains with diameters between 15 and 30 µm cemented together by clay platelets or calcium carbonate, in accordance with the grain size distribution. The subangular shape of the grains and their aspect are typical of loess (Barden *et al*.,1973; Grabowska-Olszewska, 1975; Osipov & Sokolov, 1995). Figure 2-b presents SEM observations of a damaged sample after one-million-cycle loading in an oedometer cell. Obviously, the original soil structure is destroyed, large inter-aggregates pores disappear, and soil particles are redistributed, leading to a more compact microstructure.

The soil water retention curve was determined using osmotic technique for suction control (Cui & Delage 1996, Delage *et al*. 1998), and it is shown in Figure 3. It appears that the air entry value is at a suction value close to zero, probably due to the high porosity of the loess studied.

## 3 A constitutive model for unsaturated structured soils under cyclic loading

It is assumed in this study that the soil investigated does not present any rate dependent behaviour. Cyclic loadings are thus dealt with as a succession of quasi-static states. Obviously, this assumption cannot be valid if the loading frequency is important.

### *3.1 Elastic constitutive relationship based on the damage theory of bond*

The cemented soil is considered as a mixture composed of the solid matrix and the bond (Figure 4), each one being associated to its own stress and strain state. Total volume $V_t$ of cemented soil is defined as

$$V_t = V_m + V_v + V_b \tag{1}$$

where the subscripts *m*, *v*, and *b* refer to the solid matrix, void and bond respectively.

A partition of the total stress between matrix and bond contributions is assumed as follows

$$p = p_m + p_b, \quad q = q_m + q_b \tag{2}$$

where *p* and *q* are respectively the net isotropic and deviatoric stresses in the triaxial stress space.

The bond is regarded as a brittle material. It is thus assumed that the bonding material can undergo only reversible elastic strains. Bond degradation may occur according to accumulated strains. With the definition of bond concentration $\beta = V_b / V_t$, the total elastic volumetric strain increment $d\varepsilon_p^e$ can then be derived as

$$d\varepsilon_p^e = (1-\beta)d\varepsilon_{pm}^e + \beta d\varepsilon_{pb}^e \tag{3}$$

where $\varepsilon_{pm}^e, \varepsilon_{pb}^e$ are apparent elastic volumetric strains of matrix and bond, defined over solid matrix phase (volume $V_m + V_v$) and bond phase (volume $V_b$) respectively.

Similarly, the total elastic deviatoric strain can be given as

$$d\varepsilon_q^e = (1-\beta)d\varepsilon_{qm}^e + \beta d\varepsilon_{qb}^e \tag{4}$$

In order to quantify the bond degradation and its effects on the overall behaviour, following the proposal of Vaunat & Gens (2003) and Carol *et al.* (2001), the following relations are assumed

$$d\varepsilon_{pb}^e / d\varepsilon_p^e = \chi_0 e^{L_0 - L} \quad ; \quad d\varepsilon_{qb}^e / d\varepsilon_q^e = \chi_1 e^{L_0 - L} \tag{5}$$

where $\chi_0$ and $\chi_1$ are positive scalars smaller than 1, $L_0$ is a scalar accounting for the energy threshold of damage occurrence, and *L* is the damage evolution variable, which is a function of the accumulated total strains:

$$L(\boldsymbol{\varepsilon}) = k_\alpha \xi_p + k_\beta \xi_q, \quad \xi_p = \int |d\varepsilon_p|, \quad \xi_q = \int |d\varepsilon_q| \qquad (6)$$

where $k_\alpha$, $k_\beta$ are material constants to be determined.

From above equations (3), (4), (5) and (6), the total elastic strain increments induced by the total stress can be expressed as

$$\begin{cases} d\varepsilon_p^e = \dfrac{dp}{K_m}(1-\beta) \bigg/ \left[1 - \beta\chi_0 + (1-\beta)\chi_0 \dfrac{K_{b0}}{K_m}\right] \\ d\varepsilon_q^e = \dfrac{dq}{3G_m}(1-\beta) \bigg/ \left[1 - \beta\chi_1 + (1-\beta)\chi_1 \dfrac{G_{b0}}{G_m}\right] \end{cases}, \quad L \leq L_0$$

$$\begin{cases} d\varepsilon_p^e = (C_{00}B_{01} - A_{01}C_{11})/(A_{00}B_{01} - B_{00}A_{01}) \\ d\varepsilon_q^e = (C_{00} - A_{00}d\varepsilon_p^e)/A_{01} \end{cases}, \quad L > L_0 \qquad (7)$$

and,

$$A_{00} = \left(\beta - (1-\beta)\frac{K_{b0}}{K_m}\right)\chi_0 e^{L_0 - L}\left(1 - \mathrm{sgn}(d\varepsilon_p)\varepsilon_p^e k_\alpha\right) - 1$$

$$A_{01} = -\left(\beta - (1-\beta)\frac{K_{b0}}{K_m}\right)\chi_0 e^{L_0 - L} \mathrm{sgn}(d\varepsilon_q)\varepsilon_p^e k_\beta$$

$$B_{00} = -\left(\beta - (1-\beta)\frac{G_{b0}}{G_m}\right)\chi_1 e^{L_0 - L} \mathrm{sgn}(d\varepsilon_p)\varepsilon_q^e k_\alpha$$

$$B_{01} = \left(\beta - (1-\beta)\frac{G_{b0}}{G_m}\right)\chi_1 e^{L_0 - L}\left(1 - \mathrm{sgn}(d\varepsilon_q)\varepsilon_q^e k_\beta\right) - 1$$

$$C_{00} = \frac{dp'}{K_m}(\beta - 1) + \left(\beta - (1-\beta)\frac{K_{b0}}{K_m}\right)\chi_0 e^{L_0 - L} \cdot \varepsilon_p^e \left(\mathrm{sgn}(d\varepsilon_v)k_\alpha d\varepsilon_p^p + \mathrm{sgn}(d\varepsilon_q)k_\beta d\varepsilon_q^p\right)$$

$$C_{11} = \frac{dq}{3G_m}(\beta - 1) + \left(\beta - (1-\beta)\frac{G_{b0}}{G_m}\right)\chi_1 e^{L_0 - L} \cdot \varepsilon_q^e \left(\mathrm{sgn}(d\varepsilon_v)k_\alpha d\varepsilon_p^p + \mathrm{sgn}(d\varepsilon_q)k_\beta d\varepsilon_q^p\right)$$

where $K_{b0}$, $G_{b0}$, $K_m$, $G_m$ are the bulk and shear moduli of bond and solid matrix respectively.

Moreover, as for saturated cemented soils, the deformation due to suction is assumed to be distributed between the solid matrix and bond according to the ratio of bond concentration $\beta$. Accordingly, the elastic strain increment induced by suction is derived as

$$d\varepsilon_{ps}^e = \left(\frac{1-\beta}{K_{ms}} + \frac{\beta}{K_{b0}}\right)ds \qquad (8)$$

where $K_{ms}$ is the bulk modulus of solid matrix under suction. It should be noted that the bulk modulus with respect to suction of the bonding material is assumed to be equal to that related to mean pressure. A physical interpretation of this assumption lies in the fact that the material constituting the cement is characterised by a porosity considerably finer than the macro-porosity of the loess. From this point of view, it is expected that the cement will remain fully saturated under usual suctions to which the studied loess is submitted in situ. As a consequence, because of the validity of Terzaghi's effective stress in the domain of positive suctions but saturated soils, bulk moduli with respect to mean pressure and suction are the same.

Finally, with Equation (7) and (8), the total elastic strain increments of unsaturated cemented soils can be obtained as

$$\begin{cases} d\varepsilon_p^e = d\varepsilon_{pm}^e + d\varepsilon_{pb}^e + d\varepsilon_{ps}^e \\ d\varepsilon_q^e = d\varepsilon_{qm}^e + d\varepsilon_{qb}^e \end{cases} \quad (9)$$

Thus the strain increments in bond are obtained using Equations (4) and (5) and the stress increments in bond can be derived as

$$\begin{cases} dp_b = K_{b0} d\varepsilon_{pb}^e \\ dq_b = 3G_{b0} d\varepsilon_{qb}^e \end{cases} \quad (10)$$

Furthermore, the corresponding stress increments in solid matrix can be given by Equation (2).

### 3.2 Unsaturated mechanical behaviour of studied soils

In order to investigate the mechanical behaviour of loess in unsaturated states, the loading-collapse (LC) yield curve proposed by Alonso *et al.* (1990), and the water retention curve (WRC) proposed by van Genuchten (1980) are used in this study.

LC yield surface is given in (*p*, *s*) plane as

$$\frac{p_0}{p^c} = \left(\frac{p_0^*}{p^c}\right)^{\frac{\lambda(0)-\kappa_m}{\lambda(s)-\kappa_m}} \quad (11)$$

where $p_0$, $p_0^*$ are preconsolidation stresses for a given suction *s* and for saturated conditions respectively, and $p^c$ is a reference net mean stress. The soil compression coefficient in unsaturated state is given as

$$\lambda(s) = \lambda(0)\left[(1-r_s)\exp(-\beta_s s) + r_s\right] \quad (12)$$

where $\lambda(0)$ is the soil compression coefficient in saturated states, $r_s$ is a constant related to the maximum stiffness of the soil, and $\beta_s$ is a parameter which controls the rate of increase of soil stiffness with suction.

The yield surface in triaxial stress space (*p*, *q*, *s*) can be established, as depicted in Figure 5 where $p_s$ is a variable introduced in the BBM model to describe the soil cohesion changes due to suction changes. $p_s$ is given as

$$p_s = k_s s \quad (13)$$

Note that both $p_{bc}$ and $p_s$ contribute to the apparent cohesion increase. Following the degradation of bond, $p_{bc}$ decreases gradually and the tensile stress of the soil approaches $p_s$ finally.

The water retention curve (WRC) is written as

$$S_r(s) = \left( \frac{1}{1 + (\beta_{Sr} \cdot s)^n} \right)^m \tag{14}$$

$\beta_{Sr}, n$ and $m$ are soil parameters and can be obtained by fitting the experimental curve (see Figure 3).

The particular case of constant water content situations is of interest. Under this assumption, the specific water volume $v_w$ is also a constant. According to the definition of $v_w$ (Wheeler, 1996):

$$v_w = 1 + e_w = 1 + S_r \cdot e \tag{15}$$

when $dv_w = 0$, the relationship between the increments of void ratio and degree of saturation can be deduced as follows

$$\frac{dS_r}{S_r} = -\frac{de}{e} \tag{16}$$

Combined with the water retention curve (Equation 14), the coupling relation between soil deformation and suction variation is deduced in the case of constant water content tests.

### 3.3 Plastic constitutive relationships based on bounding surface theory

The model under development is aimed at simulating the cyclic behaviour of unsaturated loessic soils. As a consequence, it appears important to include in the modelling framework some feature that enables the occurrence of irreversible strains along the cyclic loading stage even if the load cycles remain in the domain of small deformation. The choice of the bounding surface theory has been made in this paper. It is combined with the previously described frameworks for damage and unsaturated soils description.

A complete set of equations of an elastoplastic model for unsaturated structured soils under cyclic loading is now formulated.

*Choice of stress variables*

To account for the effect of bond damage on the plastic deformation of soils, a pair of stress variables $\bar{\sigma} = (\bar{p}, \bar{q})^T$ in the triaxial space is determined as below:

$$\bar{p} = p + \exp(L_0 - L)p_{bc} \; ; \; \bar{q} = q + \exp(L_0 - L)q_{bc} \tag{17}$$

Simultaneously, the hardening parameter $\bar{p}_0$ is given as

$$\bar{p}_0 = (1 + \chi)p_0, \; \chi = \chi_0 \exp(L_0 - L) \tag{18}$$

*Yield surface equation*

Here the formulation proposed by Pastor *et al*. (1985) is used. According to the fitting of experimental curves of the dilatancy coefficient and the stress ratio, the plastic potential surface equation is devised as followed:

$$G(\vec{\sigma}, p_0^*, L, s) = \bar{q} - M_g(\bar{p} + p_s)(1 + 1/\alpha_g)[1 - (\frac{\bar{p} + p_s}{\bar{p}_0 + p_s})^{\alpha_g}] \quad (19)$$

Furthermore, the corresponding yield surface can be given as

$$F(\vec{\sigma}, p_0^*, L, s) = \bar{q} - M_f(\bar{p} + p_s)(1 + 1/\alpha_f)[1 - (\frac{\bar{p} + p_s}{\bar{p}_0 + p_s})^{\alpha_f}] \quad (20)$$

where $M_g$ is the slope of the critical state line, and $\alpha_g$ is a constant related with the dilatancy coefficient. $M_f$, $\alpha_f$ are constant without definite physical meanings, but with $M_f/M_g$ associated with the relative density of soils. As in Dafalias & Herrmann (1982), the bounding surface is assumed to coincide with the yield surface.

*Non-associated flow rule*

The non-associated flow rule is assumed, and given by

$$d\vec{\varepsilon}^p = (d\varepsilon_p^p, d\varepsilon_q^p)^T = \bar{n}_{gL/U} \cdot \frac{\bar{n}_f^T d\vec{\sigma}}{H_{L/U}} \quad (21)$$

with $\bar{n}_{gL/U}$ and $\bar{n}_f$, normal vectors to respectively the plastic potential surface during loading or unloading and bounding surface, and $H_{L/U}$, the plastic modulus during loading or unloading.

*Hardening law*

As for the hardening law, with respect to one specified value of suction, the hardening parameter $p_0$ is dependent on the volumetric strain as well as the deviatoric strain simultaneously

$$dp_0 = \frac{\partial p_0}{\partial \varepsilon_m^p} \cdot d\varepsilon_{pm}^p + \frac{\partial p_0}{\partial \varepsilon_m^q} \cdot d\varepsilon_{qm}^q \quad (22)$$

and the hardening law employed can be expressed as

$$\begin{cases} \frac{\partial p_0}{\partial \varepsilon_{pm}^p} = \frac{1 + e_m}{\lambda_m - k_m} \cdot p_0 \\ \frac{\partial p_0}{\partial \varepsilon_{qm}^p} = \frac{\partial p_0}{\partial \xi_{qm}^p} \frac{\partial \xi_{qm}^p}{\partial \varepsilon_{qm}^p} = \beta_0 \beta_1 \exp(-\beta \xi_{qm}^p) \frac{\partial \xi_{qm}^p}{\partial \varepsilon_{qm}^p} \frac{\partial p_0}{\partial \varepsilon_{pm}^p} \end{cases} \quad (23)$$

where $\beta_0, \beta_1$ are the hardening coefficients, and $\xi_{qm}^p$ is the absolute accumulation of plastic deviatoric strain, as

$$\xi_{qm}^p = \int |d\varepsilon_{qm}^p| \quad (24)$$

Combined with the LC yield curve (Equation 11), the evolution of hardening parameter $p_0^*$ in saturated state is derived

$$\frac{dp_0^*}{p_0^*} = \frac{1+e_m}{\lambda(0)-\kappa_m}\left[d\varepsilon_{pm}^p + \beta_0\beta_1\exp(-\beta\xi_{qm}^p)\frac{\partial \xi_{qm}^p}{\partial \varepsilon_{qm}^p}d\varepsilon_{qm}^p\right] \qquad (25)$$

According to the consistency condition, $dF(\vec{\sigma}, p_0^*, L, s) = 0$, and with the non-associated flow rule and hardening law above, the plastic modulus on the bounding surface $H_L^{BS}$ can be obtained.

*Mapping rule during loading*

According to bounding surface theory, the plastic modulus, $H_{L/U}$, at the current stress point $\overline{P}$ is a function of the plastic modulus, $H_{L/U}^{BS}$, at the corresponding image stress point $\overline{P}_I$ on the bounding surface. This allows for plastic strains generation within the domain delimited by the bounding surface, during both loading and unloading stages. Here a radial mapping rule between the current stress point and image stress point on the bounding surface is employed, that is, amount of plastic modulus of current stress point is a function of the distance between the current point and its image.

With the consideration of the bond damage and unsaturated soil mechanics, the mapping rule is defined in the newly translated coordinate system $(\overline{p}_s, \overline{q}_s)$. As seen in Figure 6, the mapping origin during loading $\overline{P}_{OL}$ is defined as the left intersection point of bounding surface with the abscissa axis $\overline{p}_s$, and the image point is determined by the intersection between the bounding surface and the line connecting the mapping origin to the current stress point.

The specific definition of the radial mapping rule on the plastic modulus at the current stress point during loading is written in the following way:

$$H_L = H_L^{BS}\left(\frac{\delta_0}{\delta}\right)^{r_L} \qquad (26)$$

with

$$r_L = r_0\left[1+\exp(L_0 - L)\right] \qquad (27)$$

where $r_0$ is the mapping exponent of the equivalent unstructured soil during loading, and $\delta_0$ and $\delta$ are the distances between the mapping origin and respectively, the image point $\overline{P}_I$ $(\overline{p}_I, \overline{q}_I)$, and the current point $\overline{P}$ $(\overline{p}, \overline{q})$ in the $(\overline{O}, \overline{p}, \overline{q})$ stress space (see Figure 6).

*Plastic modulus during unloading*

During unloading, the plastic modulus of current stress point is assumed to be correlated with the stress ratio, $\eta_U$ of the start point during unloading, and is described as:

$$H_U = \begin{cases} H_{U0}\left(\dfrac{\eta_U}{M_g}\right)^{-r_U}, & \dfrac{\eta_U}{M_g} < 1 \\ H_{U0}, & \dfrac{\eta_U}{M_g} \geq 1 \end{cases} \qquad (28)$$

where $H_{U0}$, $r_U$ are the initial plastic modulus and exponential during unloading.

Therefore, the plastic modulus above (Equation 26 or 28) combined with the flow rule (Equation 21), gives the plastic strain increments vector $d\bar{\varepsilon}^p$ during loading and unloading.

*Loading criteria*

According to the plasticity theory, the loading criteria should be given as follows:

$$\begin{cases} \dfrac{\bar{n}_f^T d\bar{\sigma}}{H_L} > 0, & Loading \\ \dfrac{\bar{n}_f^T d\bar{\sigma}}{H_L} = 0, & Neutral\ loading \\ \dfrac{\bar{n}_f^T d\bar{\sigma}}{H_U} < 0, & Unloading \end{cases} \quad (29)$$

### *3.4 Determination of model parameters*

Most of the parameters of the proposed model can be obtained directly from laboratory experiments, or indirectly by fitting analysis of correlative test results. For instance, tests that involve isotropic drained compression ( loading and unloading ) at different constant suction values provide data to find $p_c, p_0^*, \lambda_0, \kappa_m, r, \beta_s$, tests that involve a drying-wetting cycle at a given net mean stress provide data to find $\kappa_s$, drained shear tests at different suction values provide data to find $M_g, k_s$, and tests of soil particle analysis may provide data to determine $\chi_0, \chi_1, \beta$. As a first estimate, it is assumed that $\beta$ is close to $\chi_0$ and that $\chi_1 = \chi_0$. The parameters related to the bounding surface like $\alpha_f, \alpha_g, \beta_0, \beta_1$ can be obtained by trial, as suggested by Zienkiwicz *et al*. (1985) and Pastor *et al*. (1985).

### 4 Model validation

To investigate the capability of this elastoplastic model to describe the behaviour of structured unsaturated soils under cyclic loading, cyclic triaxial tests are simulated and the numerical results compared to experimental results obtained in the laboratory. Since the natural water content in the loess profile is expected to change due to seasonal effects, the effect of initial water content on loess behaviour is investigated on the 2.2 m sample. Three different water contents are investigated more precisely: 18% (natural water content), 23% and 29%. The initial water contents were obtained in the laboratory by adding water using a wet filter paper. All samples were first consolidated with a confining stress of 25 kPa. Then the single or multi-level cyclic loadings with a frequency of 0.05 Hz were applied using a cyclic triaxial cell described by Cui *et al*. (2007).

Three samples with the above water contents (assumed to be constant during the whole tests) are first cyclically sheared to a deviator value of 15 kPa. Subsequently, several cycles of loading and unloading at different levels of deviator stress are performed. At each level, the peak deviatoric stress increases by 15 kPa and each level of loading runs 100 cycles until the sample reaches failure. The comparison of results between experiments and model predictions is depicted in Figure 7 in terms of axial stress vs. number of cycles. The model parameters used are listed in Table 1. Figure 7 also presents the effects of the account for bonding and its damage in the simulation. As expected, the model is able to reproduce volumetric strain accumulation with the load cycles. Furthermore, the consideration of bond damage in the model gives much better simulation results, especially for case (a) (w = 18%) and case (b) (w = 23%).

## 5 Conclusion

In this paper, a model for describing the mechanical and hydraulic behaviour of cemented unsaturated loess is presented. This soil is a typical homogeneous yellowish-grey, porous calcareous loess, mainly composed of quartz and feldspar with some clay. SEM observations show the presence of large aggregates with associated inter-aggregate pores. The presence of clay platelets and calcium carbonate may act as bonds to cement the solid grains. This soil is characterized by low plasticity, low natural degree of saturation, low clay fraction, and relatively high calcium carbonate content.

On the basis of the bounding surface model, damage theory and unsaturated soil mechanics, an elastoplastic model which includes structure damage for unsaturated loess under cyclic loading has been elaborated. The chosen law for bond degradation links structure damage to the accumulation of strain. The BBM model was considered for the suction effect.

Multi-level cyclic triaxial tests were simulated to verify the predictions of the model. Different water contents were considered. The outcome is encouraging as the model seems to be able to predict the behaviour of unsaturated cemented soils under cyclic loading.

The current constitutive model for unsaturated cemented soils is still complicated, with many parameters needed to be determined by various experiments. In the future, focus should be put on the simplification of the cement damage part of this model. Furthermore, the bounding surface theory should be modified to reflect the hysteretic behaviour of soils under cyclic loading.

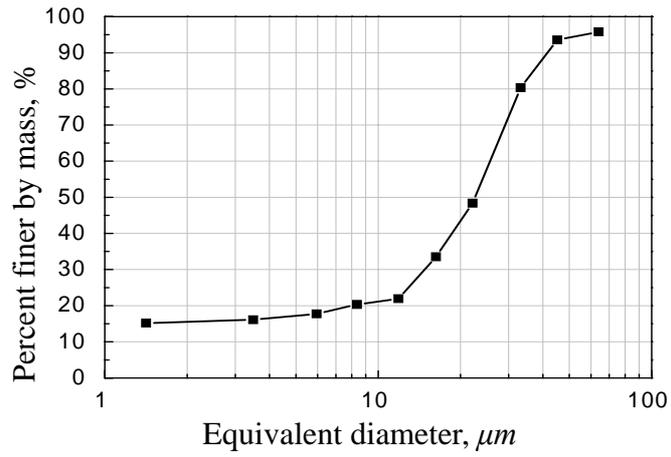

Figure 1. Grain size distribution curve of the studied loess.

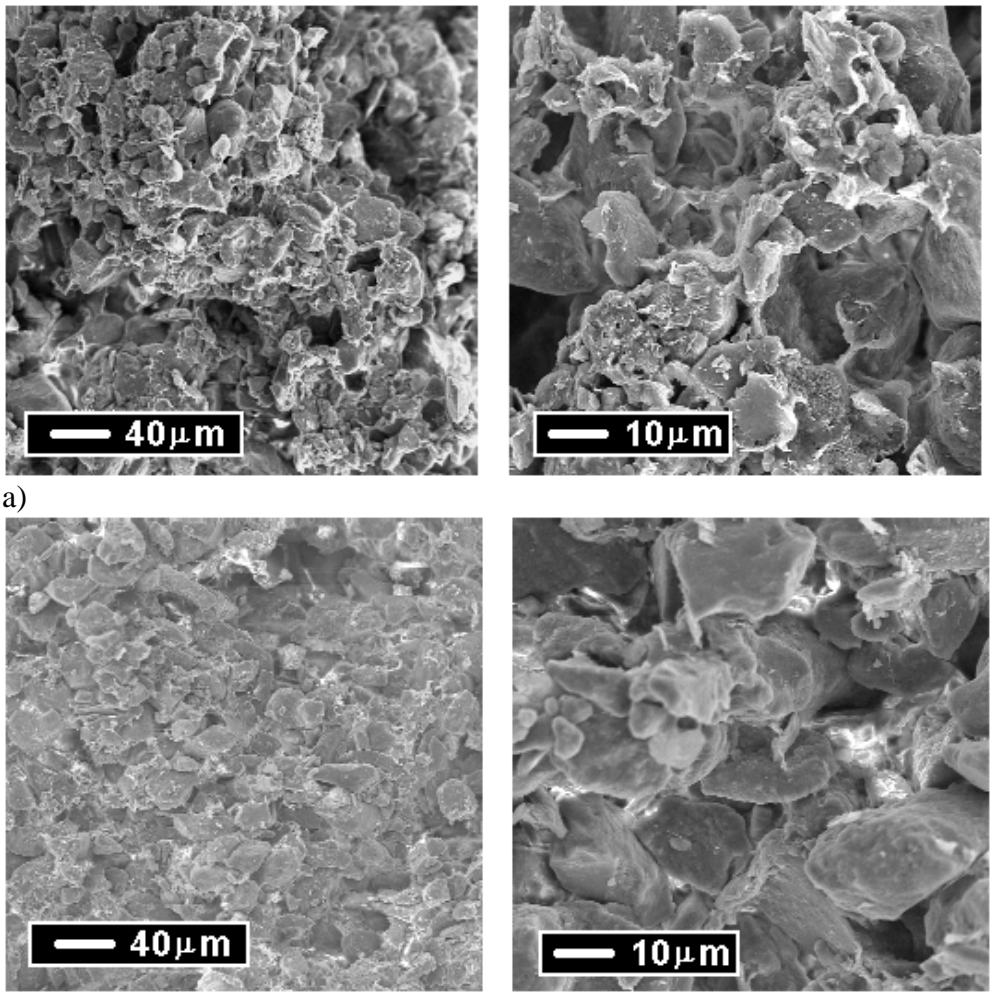

a)

b)

Figure 2. SEM observation of samples from a depth of 2.2m; a) Intact sample; b) Damaged sample.

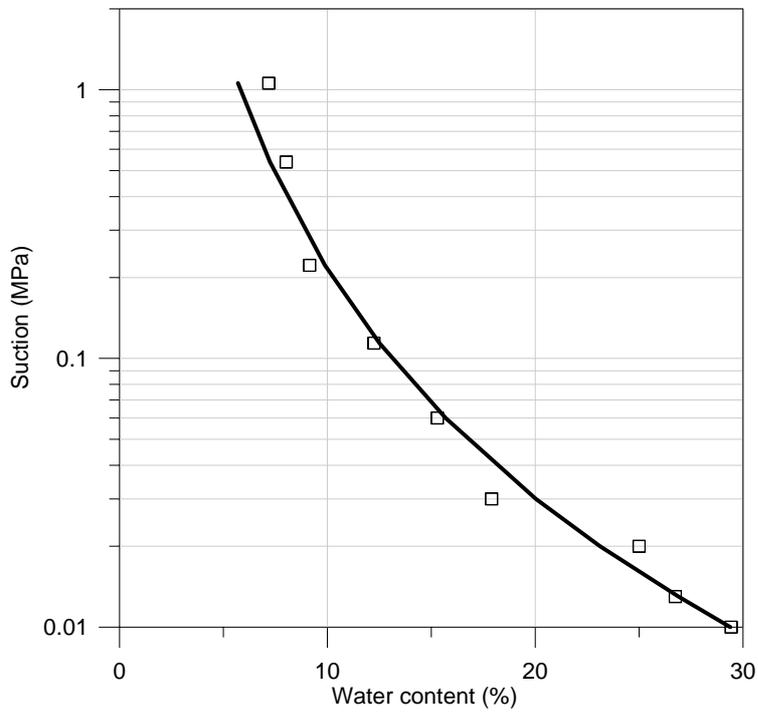

**Figure 3. Water retention curve of the studied loess: experimental curve (symbols) and curve fitted using van Genuchten model (continuous line).**

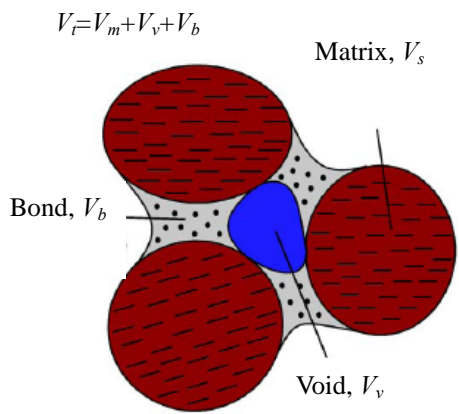

Figure 4. Schematic arrangement of soil structure (after Garitte *et al*., 2006)

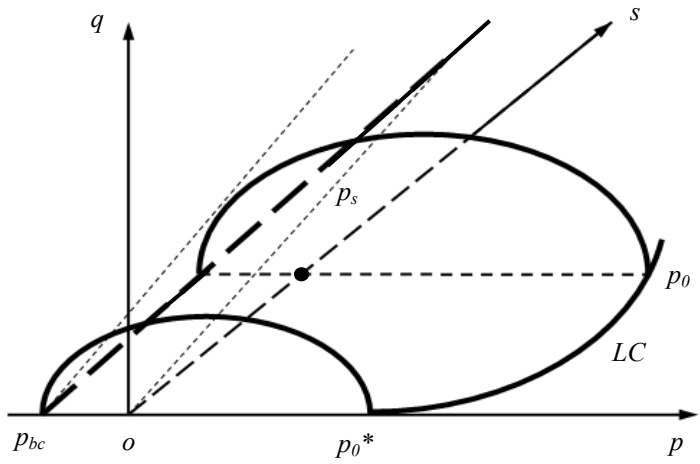

Figure 5. Three-dimensional view of yield surfaces in (*p*,*q*,*s*) stress space

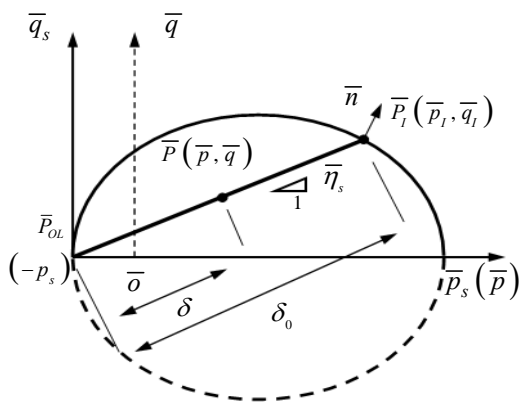

**Figure 6. Mapping rule in bounding surface model**

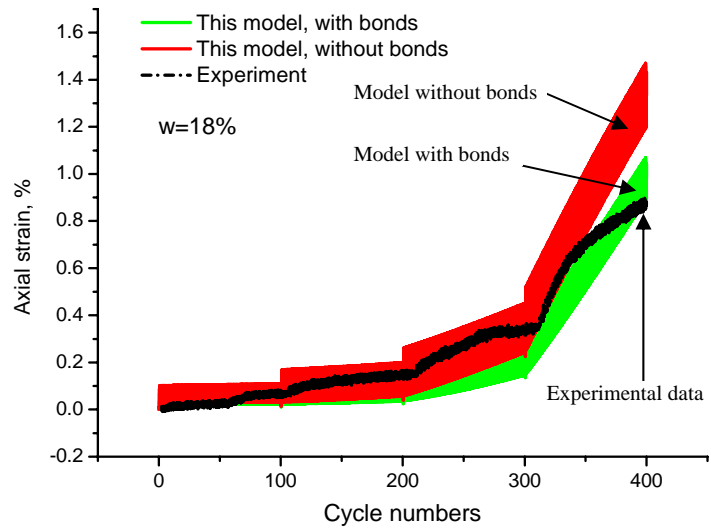

a)

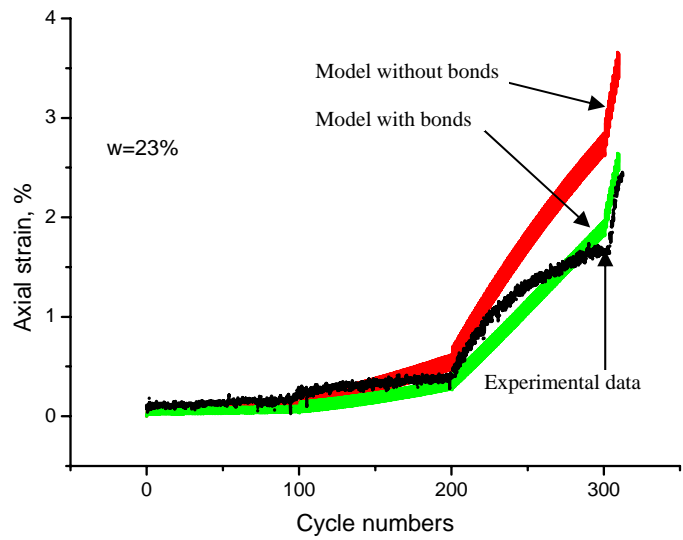

b)

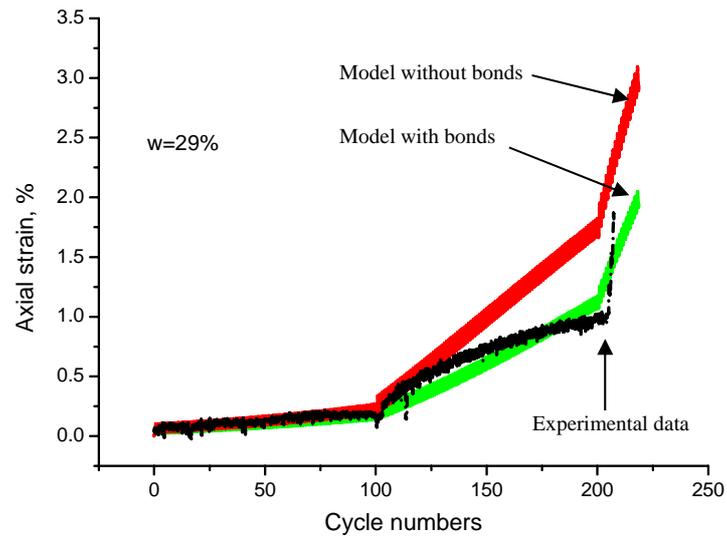

c)

**Figure 7. Multi-level cyclic triaxial tests for samples with 3 different water contents: a) w=23%; b) w=18%; c) w=29%**

Table 1. Parameters used in simulation of the cyclic triaxial shear tests

| $e_0$ | $\lambda_0$ | $\kappa_m$ | $u_m$ | $M_g$ | $M_f$ | $\alpha_g$ | $\alpha_f$ | $\beta_0$ | $\beta_1$ |
|---|---|---|---|---|---|---|---|---|---|
| 0.93 | 0.17 | 0.012 | 0.25 | 1.35 | 0.6 | 0.45 | 0.45 | 4.30 | 0.23 |
| $r_L$ | $r_U$ | $\chi_0=\chi_1$ | $\beta$ | $k_\alpha=k_\beta$ | $u_b$ | $r_s$ | $\beta_s$ | $k_s$ | $\kappa_s$ |
| 1.4 | 1.85 | 0.35 | 0.35 | 2.0 | 0.25 | 0.75 | 0.01 | 0.02 | 0.01 |
| $K_b$ (kPa) | $p_{bc}$ (kPa) | $q_{bc}$ (kPa) | $s_0$ (kPa) | $p^c$ (kPa) | $p_0^*$ (kPa) | $H_U$ (MPa) | $n$ | $m$ | $\beta_{Sr}$ |
| 5000 | 10.0 | 0.0 | 1000 | 25 | 1000 | 50000 | 5.75 | 0.06 | 154.76 |